# Ternary inorganic compounds containing carbon, nitrogen, and oxygen at high pressures

Brad A. Steele and Ivan I. Oleynik*

*Department of Physics, University of South Florida, Tampa, FL 33620*

E-mail: oleynik@usf.edu

## Abstract

Ternary $C_xN_yO_z$ compounds are actively researched as novel high energy density and ultrahard materials. Although some synthesis work has been performed at ambient conditions, very little is known about the high pressure chemistry of of $C_xN_yO_z$ compounds. In this work, first principles variable-composition evolutionary structure prediction calculations are performed with the goal of discovering novel mixed $C_xN_yO_z$ materials at ambient and high pressure conditions. By systematically searching ternary variable composition crystalline materials, the full ternary phase diagram is constructed in the range of pressures from 0 to 100 GPa. The search finds the $C_2N_2O$ crystal containing extended covalent network of C, N, and O atoms, having space group symmetry $Cmc2_1$, and stable above just 10 GPa. Several other novel metastable $(CO)_x$-$(N)_y$ crystalline compounds discovered during the search, including two polymorphs of $C_2NO_2$ and two polymorphs of $C_3N_2O_3$ crystals are found to be energetically favorable compared to polymeric carbon monoxide (CO) and nitrogen. Predicted new compounds are characterized by their Raman spectra and equations of state.

# Introduction

Carbon, nitrogen, and oxygen are among the most abundant elements on the planet, yet very little is known about the condensed phases consisting of all three elements only. Numerous metastable CNO compounds have been studied at ambient conditions[1–4] but they are not thermodynamically stable and undergo exothermic reactions to release $N_2$, $O_2$, $CO_2$, and other gases depending on the stoichiometry. These compounds as well as carbon monoxide (CO) become polymeric at high pressures due to dramatic changes in the chemical bonding. Such changes might be exploited to stabilize novel mixed $C_xN_yO_z$ compounds, both in molecular as well as extended covalent network forms of crystalline compounds. Nitrogen transforms from a triple bonded molecular crystal into a polymeric single bonded crystal at pressures near 100 GPa[5,6], carbon dioxide transforms from a double bonded molecular crystal into an extended six-fold coordinated crystal above 50 GPa[7], and oxygen transforms into a metal above 96 GPa[8]. In contrast, carbon monoxide transforms into an extended polymeric amorphous material at relatively low pressure just above 5 GPa[9].

Triply-bonded polymeric cubic gauche nitrogen (cg-N) would be an excellent energetic material as it releases enormous amount of energy upon conversion of single bonds in Cg-N to triple bonds in $N_2$[10]. Although Cg-N is stable at high pressures above 100 GPa[5,11], its recovery at ambient conditions seems to be impossible. Under pressure, both carbon monoxide and nitrogen undergo polymerization. However, there exists a large difference in pressure at which they become polymeric: cg-N polymerizes slightly over100 GPa[5,6], whereas CO – slightly over 5 GPa[7]. By mixing together both structural units, it might be possible to introduce additional covalent bonding between C, N, and O atoms resulting in a reduced pressure of polymerization or even the formation of extended solids at ambient conditions.

Raza *et al* theoretically predicted several polymeric C, N, O crystals with stoichiometries CNO and $CN_2O$ in a pressure range of 10-100 GPa[12]. Additional investigation ofthe $CN_2O$ stoichiometry revealed similar structures[13]. It was found that CNO would decompose exothermically into $\alpha$-$C_3N_4$, $CO_2$-*I$\bar{4}$2d*, and cg-N at 60 GPa[12], while if it existed at 0 GPa, it would release 2.2 kJ/g of energy upon decomposition. Although these compounds are not thermodynamically stable, the predictions indicate that high pressures can facilitate chemical bonding between these elements to form metastable polymeric crystals at high pressures.

High pressures also stabilize a wide variety all-nitrogen- single or double bonded compounds that consist of energetic $N_5$ rings, $N_6$ rings, short $N_4$ chains, and infinite N chains[14–19]. These results and the potential applications of poly-nitrogens motivate the search for such new compounds.

In this work, the high pressure ternary $C_xN_yO_z$ crystal structures of carbon, nitrogen, and oxygen are systematically investigated using first-principles structure prediction method with the ultimate goal of finding all stable and metastable crystals and constructing the ternary C-N-O phase diagram for crystalline materials consisting of C, N, and O elements. The thermodynamic stability of the predicted structures is assessed in respect to the stable single element and any binary phases of C, N, and O as well as to metastable poly-CO and cg-N phases, which allows us to reveal new metastable $C_xN_yO_z$ compounds. The chemical bonding in these structures is analyzed by calculating the charges on the atoms and the bond orders. Characterization of the materials is performed by calculating the pressure-volume equations of state and Raman spectra.

## Computational Details

New ternary $C_xN_yO_z$ compounds with variable stoichiometry are searched at 20 GPa and 100 GPa using the first principles evolutionary structure prediction method USPEX[20–22]. At each pressure, two independent ternary searches are performed with 4-11 atoms/unit cell and 12-16 atoms/unit cell to increase the probability of finding structures with nontrivial composition and chemical bonding. Due to the complexity of the ternary search, binary structure searches are also performed with stoichiometries varying between carbon monoxide and nitrogen, $(CO)_x+N_y$, as well as between carbon nitride and oxygen; $(CN)_x+O_y$. The binary searches use 8-20 atoms/unit cell. By ensuring the binary and ternary structure searches find the same structures, we prove that complex ternary search covers the entire chemical lanscape and does not miss any potentially important stoichiometries. After the ternary and binary variable composition searches are completed, fixed composition structure searches are performed for stoichiometries on the convex hull and near the convex hull with larger number of atoms (up to 40 atoms/unit cell) to find the lowest enthalpy phases for each stoichiometry.

The construction of the ternary convex hull requires the knowledge of the stable phases of pure elemental compounds at the corners and stable binary compounds on the boundaries of the convex hull. Therefore, the following phases labeled by space group symmetry are used: R-3c and I$2_1$3 for $N_2$[11,23], Cmcm for $O_2$[8,24], Fd-3m for C[25], P$2_1$/c for $NO_2$, P-1 and C2/c for $N_2O_5$[26], P$4_2$/mnm and I-42d for $CO_2$[7,27], P31c for $C_3N_4$[28], P$2_1$ for $CN_3$[28], and P$4_2$/m and Pnnm for CN[28], corresponding phases being used at a specific pressure of their stability.

During the structure search, the cell parameters and the atomic coordinates of each structure are varied to minimize the total enthalpy at a given pressure using the Perdew-Burke-Ernzerhof (PBE) generalized gradient approximation (GGA)[29] within density functional theory (DFT) as implemented in VASP[30]. For the structure search, projector augmented wave (PAW) pseudopotentials[31] and plane wave basis sets are used with an energy cutoff of 470 eV and a 0.07 Å$^{-1}$ k-point sampling. After the structure search is completed, more accurate calculations are performed to compute the convex hulls using hard PAW pseudopotentials for N with inner core radii of 0.582 Å for N, 0.794 Å for C, and 0.804 Å for O, and energy cutoff of 1,000 eV and 0.05 Å$^{-1}$ k-point sampling. Charges on atoms and bond orders are calculated using LCAO code DMol[32].

The vibrational spectrum is computed using the frozen phonon method, which calculates harmonic frequencies by diagonalizing the dynamical matrix obtained from finite differencing of the atomic forces in respect to small atomic displacements. The off-resonant Raman intensities are computed 300 K and laser wavelength of 632.8 nm using derivatives of atomic polarizabilities in respect to applied electric fields within a linear response framework[33].

## Results and Discussion

The ternary $C_xN_yO_z$ structure search found one material on the formation enthalpy versus ternary composition convex hull that does not lie on its boundaries: the $C_2N_2O$-Cmc$2_1$ crystal with stoichiometry $C_2N_2O$ and space group Cmc$2_1$, which is stable at both 20 and 100 GPa, see Figure 1(a) and Figure 1(b). Its crystal structure is shown in Figure 2(a). The structure of this crystal is analogous to a known crystal structure of $Si_2N_2O$ and has also been investigated theoretically at ambient conditions by following this analogy and replacing Si with C[34,35]. Our calculations show it is thermodynamically stable from 10-100 GPa and does not have imaginary bands in the phonon dispersion curves at 0 GPa[35]. The phonon dispersion curves rely on the harmonic expansion of the energy, however at higher temperatures anharmonic terms in the expansion may become relevant. To properly accound for the anharmonic effects, density functional theory molecular dynamics (DFTMD) simulationsare performed at 4,000 K and 10 GPa that indicate $C_2N_2O$-Cmc$2_1$ is dynamically stable at high pressures, see Supplemental Figure S1. Previous attempts to synthesize crystalline-phase $C_2N_2O$ at high pressures by compressing gas-phase polyisocyanate $C_2N_2O$[36] were unsuccessful as the precursor transforms to an amorphous polymer[37].

A color scheme in Figs. 1(a) and 1(b) is used to show the distance in eV/atom to the ternary convex hull in Figure 1(a) and Figure 1(b). The lighter the color, the further away from the convex hull, and hence the less stable the crystal structure is. Polymeric CO is metastable at 20 GPa, but our calculations show poly-CO with space group $I2_12_12_1$[38] is 177 meV/atom from the hull. It is therefore conceivable that any structure closer to the hull than about 177 meV/atom is potentially metastable. Therefore, we explore the novel structures with formation enthalpy within this interval from the convex hull.

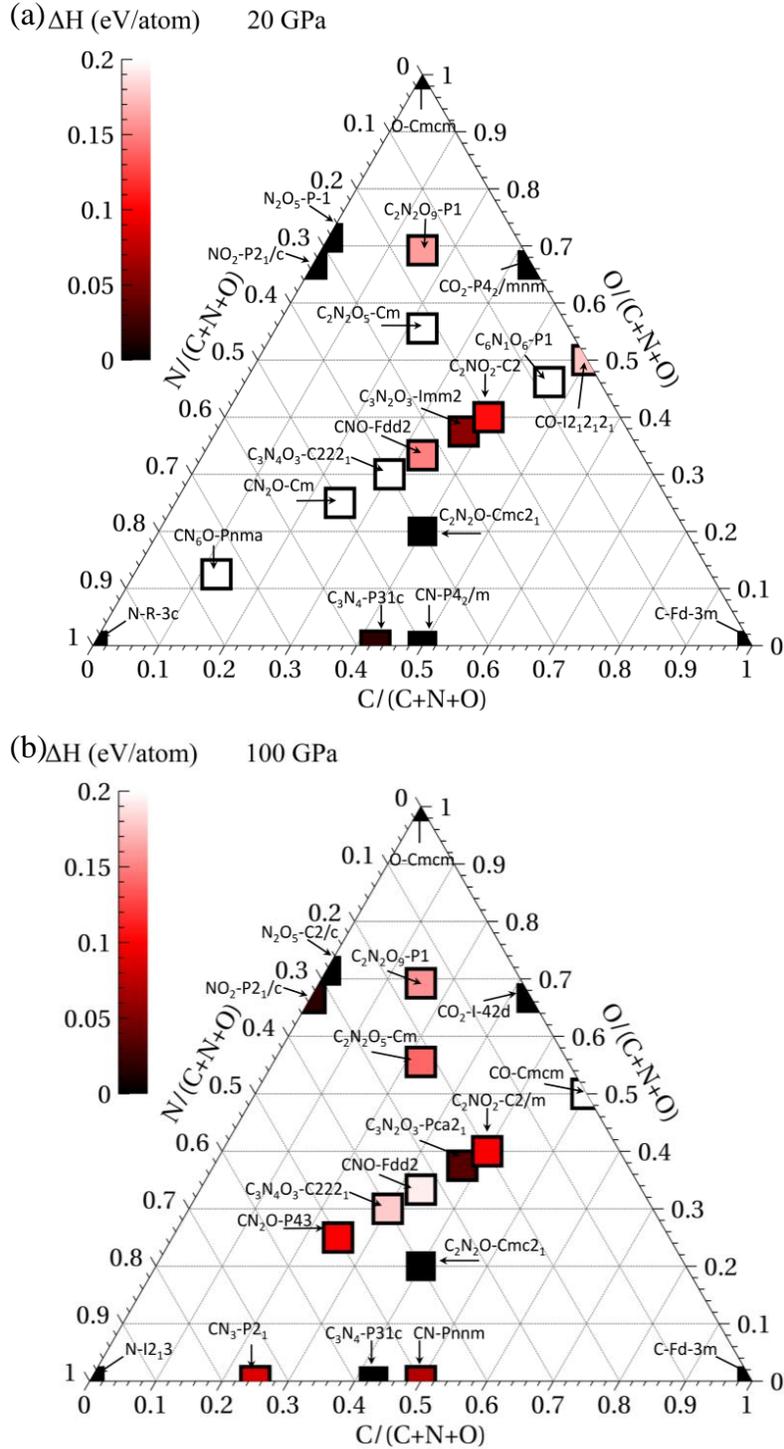

Figure 1: Ternary composition-formation enthalpy phase diagram of the $C_xN_yO_z$ compounds at (a) 20 GPa and (b) 100 GPa. Each structure is colored by the distance from the convex hull in eV/atom, see the color bar legend on the left. Structures with lower (negative) formation enthalpy are closer to the convex hull and therefore colored by more darker color.

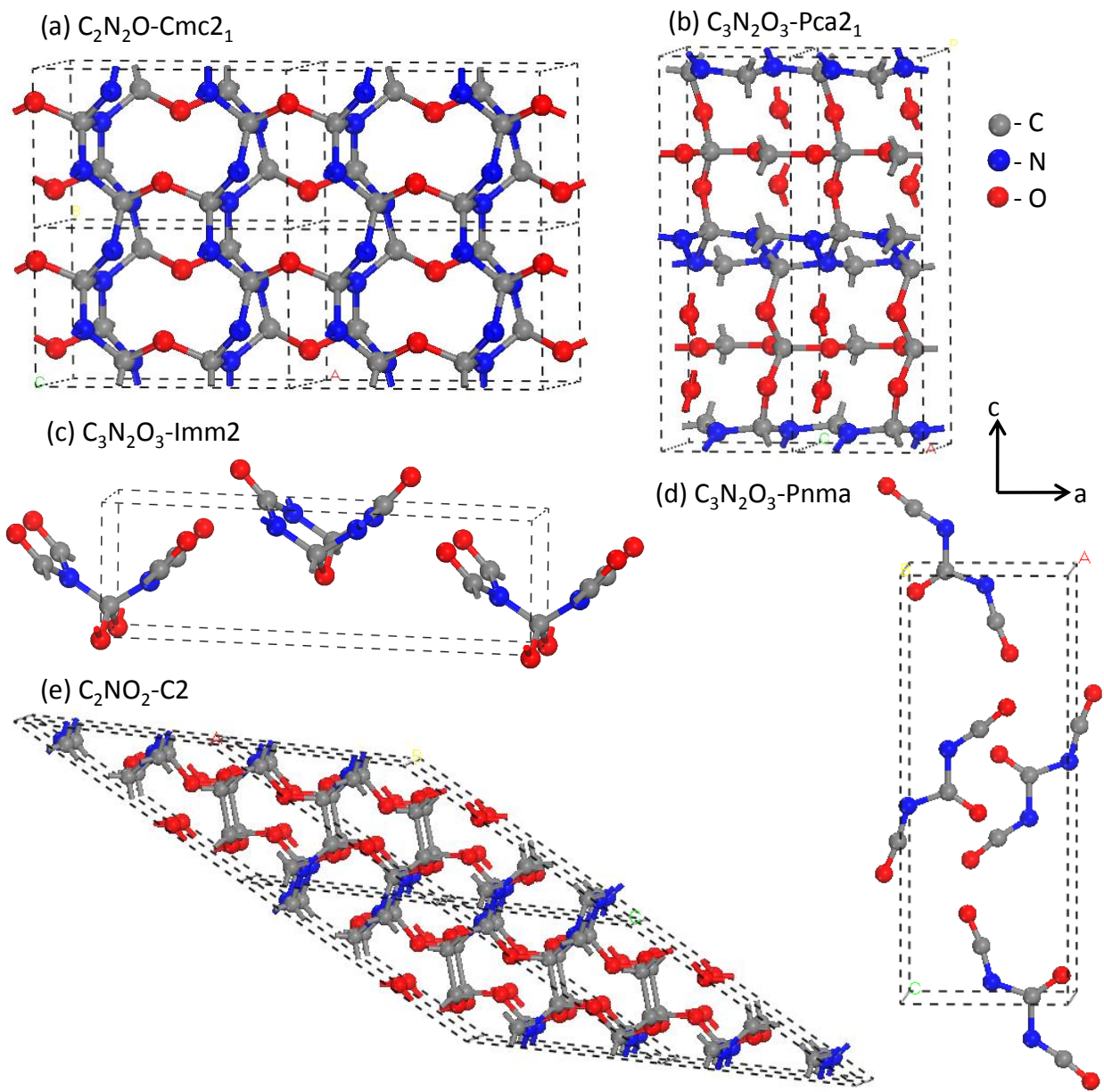

Figure 2: The crystal structure of (a) $C_2N_2O$-$Cmc2_1$, (b) $C_3N_2O_3$-$Pca2_1$, (c) $C_3N_2O_3$-$Imm2$, (d) $C_3N_2O_3$-$Pnma$[1], (e) $C_2NO_2$-$C2$. All structures displayed in the figure are at 20 GPa except of one - $C_3N_2O_3$-$Pnma$ which is at 0 GPa.

The next closest crystal structure to the ternary convex hull has stoichiometry $C_3N_2O_3$ with space group Pca2$_1$ and it has three-dimensional polymeric structure as shown in Figure 2(b). For this compound, the distance to the hull is reduced significantly at pressures from 20 GPa to 100 GPa and is only about 30 meV/atom from the convex hull at 100 GPa, which indicates it may be on the hull, i.e. thermodynamically stable, at higher pressures. DFT MD simulations at 1,000 K and 20 GPa show this material is dynamically stable at high pressures, see Supplemental Figure S2.

At lower pressures another compound close to the convex hull is found with stoichiometry $C_3N_2O_3$ and space group Imm2. It has one-dimensional polymeric structure, each unit cell containing two polymeric $C_3N_2O_3$ chains, see Figure 2(c). Recently, Klapotke *et al* synthesized a similar compound with the same stoichiometry, carbonyl diisocyanate $C_3N_2O_3$, its crystal structure with the space group Pnma consisting of individual carbonyl diisocyanate $CO(NCO)_2$ molecules weakly interacting by van der Waals forces, see Figure 2(d)[1]. The relative enthalpy between these three crystals, $C_3N_2O_3$-Pnma, $C_3N_2O_3$-Imm2 and $C_3N_2O_3$-Pca2$_1$ is plotted as a function of pressures in Figure 3. The ambient conditions $C_3N_2O_3$-Pnma crystal is the lowest enthalpy structure at pressures up to 3.8 GPa, above which the $C_3N_2O_3$-Imm2 is energetically preferred, followed by $C_3N_2O_3$-Pca2$_1$ structure, which is the lowest enthalpy structure above 17.3 GPa, see Figure 3. As our results indicate, both $C_3N_2O_3$-Imm2 and $C_3N_2O_3$-Pca2$_1$ are real candidates for high pressure synthesis.

Another highly covalent crystal structure found during the search with the formation enthalpy close to the convex hull is the one with stoichiometry $C_2NO_2$ and space group C2, shown in Figure 2(e). DFT MD simulations at 2,000 K and 20 GPa also show this material is dynamically stable at high pressures, see Supplemental Figure S3. A different polymorph of this stoichiometry with the space group symmetry C2/m was also found, but $C_2NO_2$-C2 is lower in formation enthalpy between 3.1 GPa and 93.9 GPa, see Figure 4. Other stoichiometries such as the ones proposed by Raza and Pickard[12](CNO and $CN_2O$) are found to have substantially high formation enthalpies (the distance to the hull at 20

GPa: CNO - 150 meV/atom, and $CN_2O$ - 208 meV/atom), see Figure 1(a), thus making their high pressure synthesis less probable.

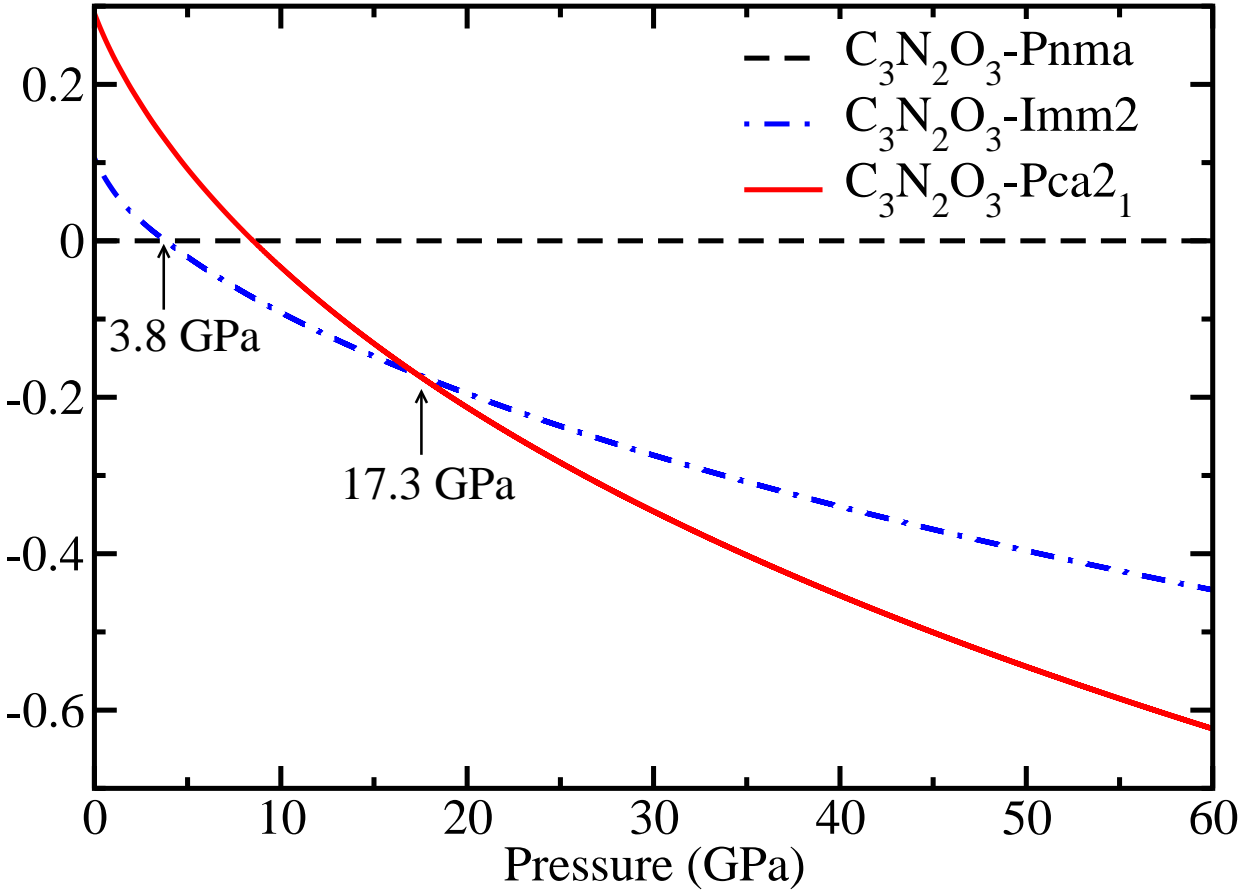

Figure 3: Relative enthalpy difference as a function of pressure between three crystal structures with stoichiometry $C_3N_2O_3$ found during the search: $C_3N_2O_3$-Pnma, $C_3N_2O_3$-Imm2 and $C_3N_2O_3$-Pca$2_1$. The reference structure is $C_3N_2O_3$-Pnma - the one synthesized by Klapotke *et al*[1].

A direct way to experimentally synthesize $C_xN_yO_z$ compounds at high pressures is to compress a mixture of CO and $N_2$ gases; such effort has been attempted recently[39,40]. Although the $C_2N_2O$-Cmc$2_1$ crystal is the only $C_xN_yO_z$ compound to be thermodynamically stable at high pressures, other metastable compounds may be accessible due to pressure and temperature activated processes during the synthesis. The reaction $2(CO)+N_2 \rightarrow C_2N_2O+O$ to make the $C_2N_2O$ crystal would require precipitation of oxygen out of the reaction volume. However, metastable compounds with stoichiometry $(CO)_xN_y$ do not require these types of processes to occur, therefore, it is expected that they might be synthesized more easily. In order to find metastable structures with stoichiometry $(CO)_xN_y$ additional structure searches are performed. The results are presented in the form of binary $(CO)_xN_y$ convex hulls in Figure 5.

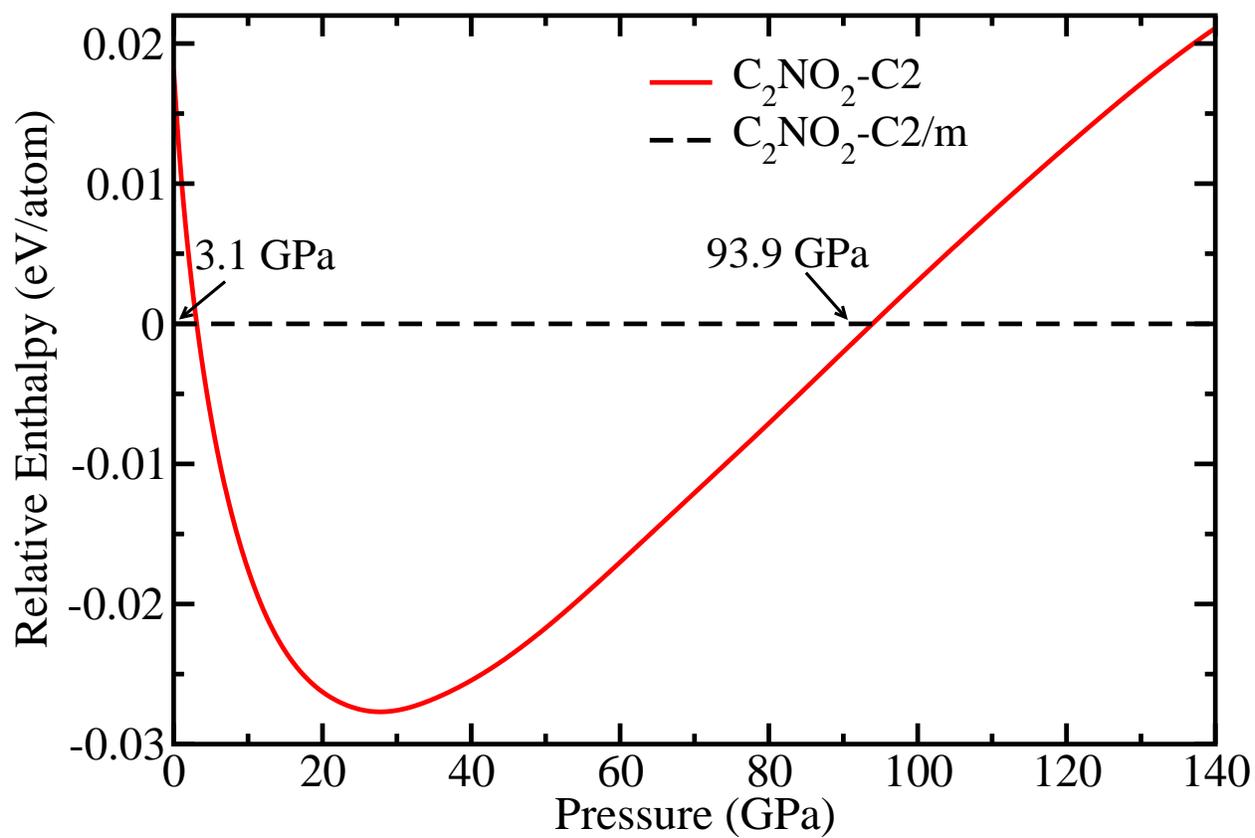

Figure 4: Relative enthalpy difference between two crystal structures with $C_2NO_2$ stoichiometry, $C_2NO_2$-C2 and $C_2NO_2$-C2/m as a function of pressure.

The most energetically favorable of these structures found during this binary search are included in the ternary hull in Figs. 1(a) and 1(b). The structures with stoichiometry $C_3N_2O_3$ and $C_2NO_2$ discussed previously are also found to be on the binary hull (Figure 5), while two other stoichiometries, CNO and $CN_2O$ proposed by Raza and Pickard are also found to have negative formation enthalpy but they are not on the binary hull neither at 20 GPa or 50 GPa. Only at 100 GPa, $CN_2O$ appears the binary hull at 100 GPa. Additional high enthalpy nitrogen-rich structures are also displayed on the binary hull, they were discussed in a recent publication[39]. These include carbonyl dipentazolate ($CN_{10}O$), carbonyl azide-pentazolate ($CN_8O$), carbonyl diazide $(CO)2N_3$[41], and a carbonyl nitrogen chain structure ($CN_4O$), see Figure 5. These structures have positive formation enthalpy at all pressures investigated making them thermodynamically unstable.

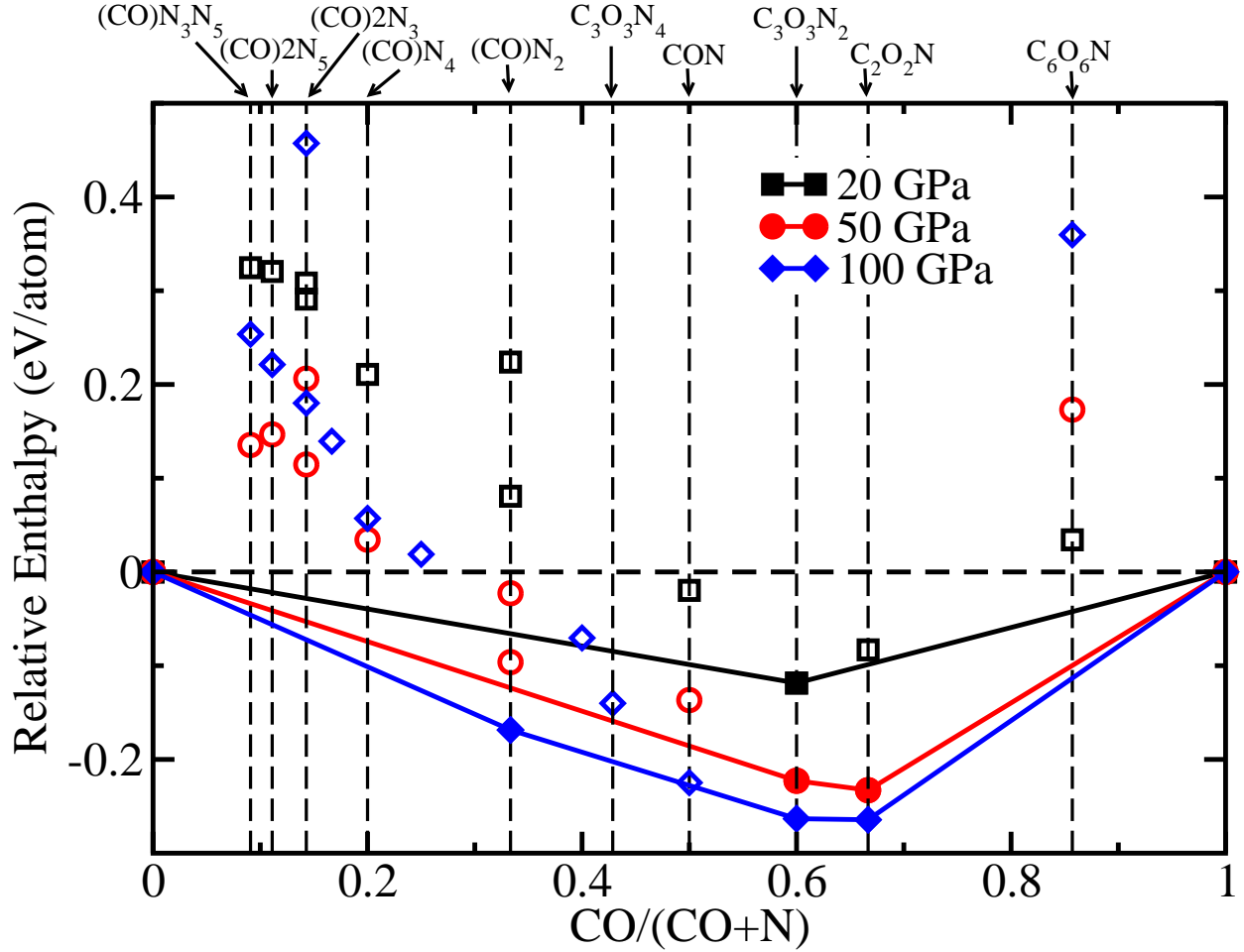

Figure 5: Binary hull for $(CO)_x N_y$ compounds at 20 GPa, 50 and 100 GPa. Stable phases relative to $CO+N_2$ are represented by solid symbols and metastable phases – by open symbols.

In terms of chemical bonding, every crystal structure found during the search, contains carbon atoms connected to their nearest neighbors by either four single bonds (coordination number 4), or two single bonds plus one double bond (coordination number 3); nitrogen atoms - by 3 single bonds (coordination number 3); and oxygen atoms – either by two single bonds (coordination number 2) or one double bond (coordination number 1). As all the atoms in each compound possess zero formal charge, they all satisfy the Pauling electroneutrality principle[42], which is used to identify compounds with stable chemical bonding.

Each structure displays a significant charge transfer from carbon to nitrogen/oxygen atoms, see Figure 6, which shows the Mulliken charges, Mayer bond orders, and bond lengths.

Some bond lengths are not given in Figure 6, but can be easily measured using the cif files provided in Supplemental Information. The carbon atoms carry a positive charge that ranges between +0.55–0.87 for the four crystal structures given in Figure 6, while nitrogen and oxygen carry a negative charge that ranges between –0.33 and –0.45. The C-N and C-O bond orders are all approximately equal to one which indicates that each bond is a single bond as to be expected from bond formation between 3-fold coorindated nitrogen and 2-fold coordinated oxygen atoms, see Figure 6. The exception is $C_3N_2O_3$-Imm2 crystal, which contains a 1-fold coodinated oxygen atom with a bond order of 1.87 indicating approximately a double C-O bond. Interestingly, nitrogen never bonds with oxygen in these structures likely due to a substantial charge transfer from carbon to nitrogen/oxygen, resulting in an electrostatic repulsion between negative N and O charges, thus making N–O bonds energetically less favorable. The C-N bond lengths are all around 1.42-1.47 Å while the C-O bond lengths are between 1.39 Å and 1.47 Å, see Figure 6. In the case of $C_3N_2O_3$-Imm2 crystal, the C-O bond length is 1.20 Å, i.e it is a double bond. $C_2NO_2$-C2 crystal contains C-C bonds with a rather large bond length of 1.70 Å and correspondingly lower bond order (0.82) due to the fact that both carbon atoms carry positive charges, resulting in electrostatic repulsion as well.

The bulk modulus of each structure is smaller than that of diamond, the largest bulk modulus being 232.5 GPa for $C_2N_2O$-Cmc$2_1$ crystal, whereas the bulk modulus for the other structures being between 200-232 GPa, see Figures 7(a),7(b), and 7(c), which is comparable with the bulk modulus of typical ceramic materials such as MgO[43,44]. The bulk modulus is calculated by fitting the calculated volume-pressure points by a 3rd order Birch-Murnaghan equation of state. The data and the fitted lines are displayed in Figures 7(a),7(b), and 7(c) together with the other parameters of the fit. Interestingly, the bulk modulus $B$ of the $C_2NO_2$-C2/m crystal is only 26.3 GPa at 0 GPa, but its pressure derivitive $B'$ is larger than that of the $C_2NO_2$-C2 crystal. Therefore, at higher pressures the bulk modulus of $C_2NO_2$-C2/m crystal is almost the same as that of $C_2NO_2$-C2. This makes the $C_2NO_2$-C2/m the denser phase at higher pressures, that is why it is the energetically favorable phase at 0 GPa and above 93.9 GPa, see Figure 4.

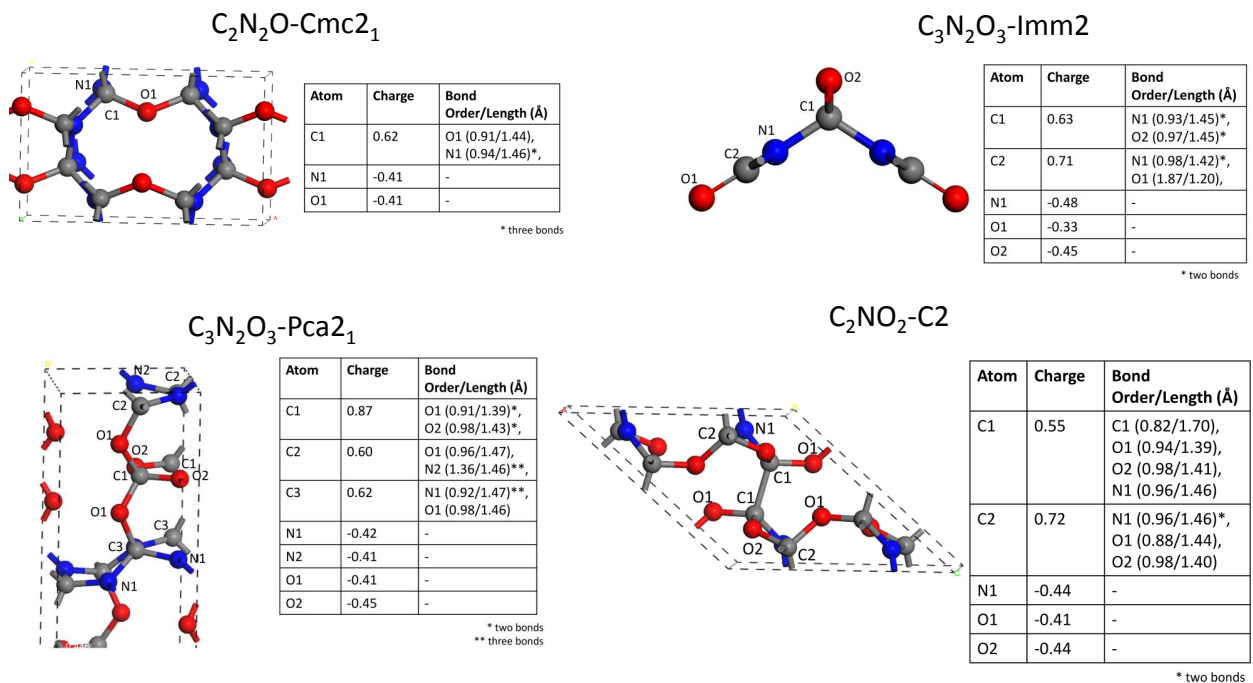

Figure 6: Mulliken charges, Mayer bond orders, and bond lengths for high-pressure $C_xN_yO_z$ crystal structures at 0 GPa.

To assist in future experimental synthetic efforts, the Raman spectra of newly predicted compounds at 20 GPa are calculated and shown in Figure 8. Due to the chemical similarity between all the structures, their Raman spectra are quite similar. The Raman spectra at 20 GPa are given in Figures 8(a),8(b), 8(c), and 8(d). Each spectrum contains C-N deformation (def.) modes in the range of frequencies between 1,000 and 1,400 cm$^{-1}$. The structures also contain C-O deformation modes in a range of frequencies between 600-1,000 cm$^{-1}$ that are in concert with a slight C-N deformation. Below 600 cm$^{-1}$ there are lattice (T) modes and several other types of modes such as C-N-C bending, see Figure 8(b); N-C-N scissoring, see Figure 8(c); CO$_3$ twisting, see Figure 8(d); and other more complicated modes involving simultaneous displacements of many atoms. Since C$_3$N$_2$O$_3$-Imm2 crystal contains 3-fold coordinated carbon atoms it has a C-N stretching mode near 1,900 cm$^{-1}$ that none of the other structures have, see Figure 8(c). However, $C_3N_2O_3$-Imm2 crystal still contains C-N deformation modes near 1,200 cm$^{-1}$ from the central 4-fold coordinated carbon atom, see Figure 2(c).

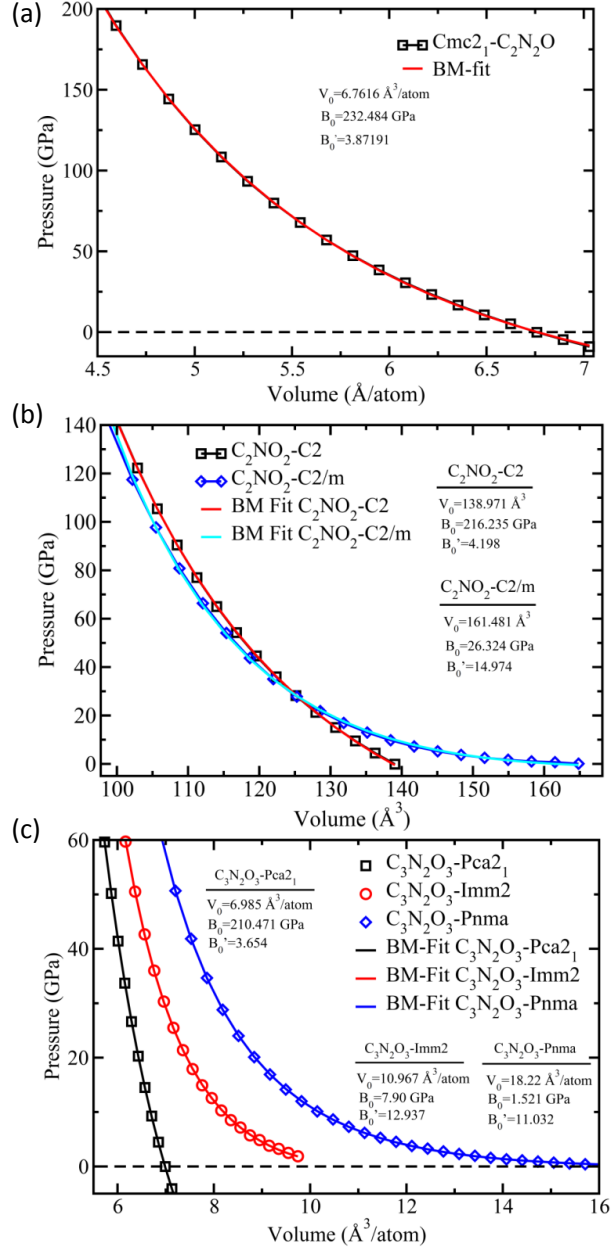

Figure 7: Pressure-volume equation of state of (a) $C_2N_2O$-$Cmc2_1$ crystal , (b) $C_2NO_2$ crystals, and (c) $C_3N_2O_3$ crystals . The squares are calculated points, and the lines – 3rd order Birch–Murnaghan fits of the the data.

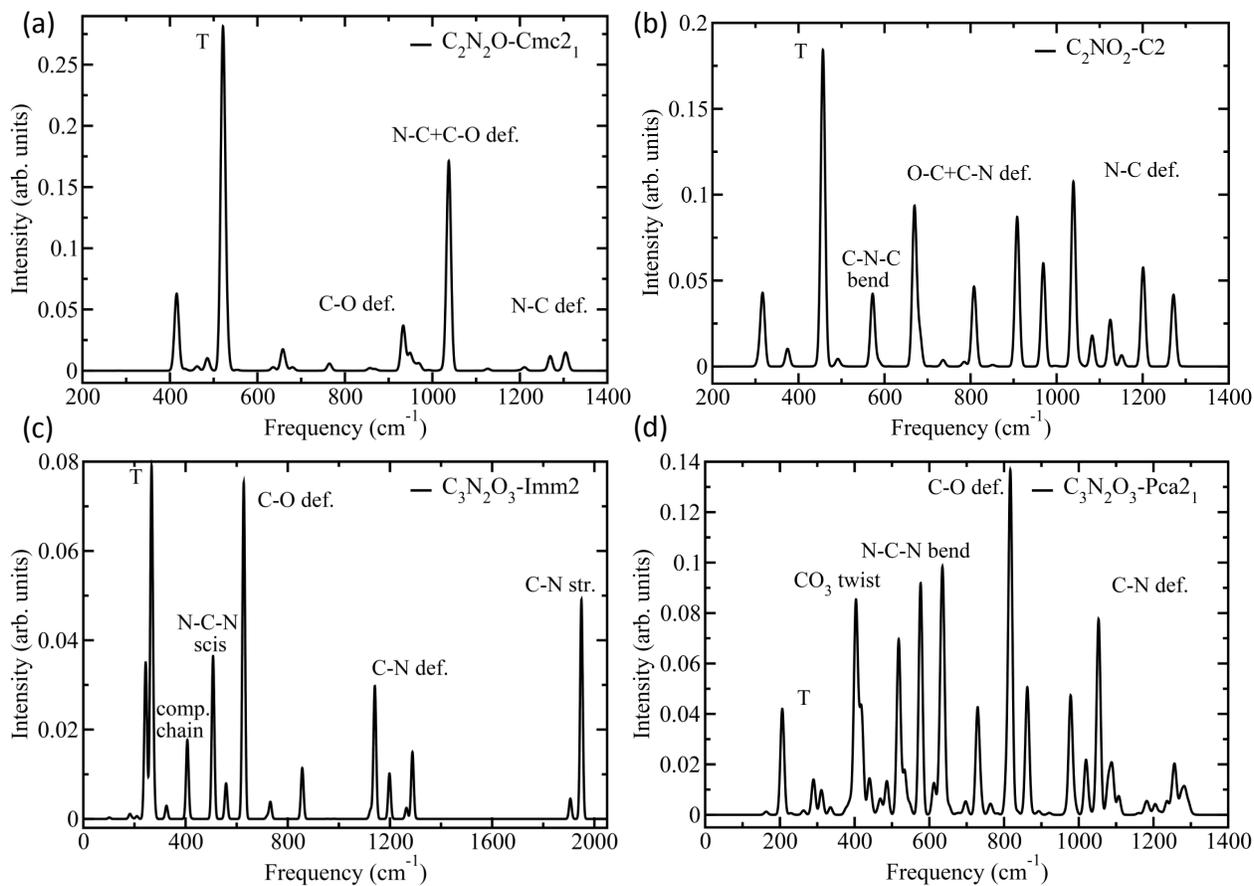

Figure 8: Raman spectrum of the (a) $C_2N_2O$-$Cmc2_1$, (b) $C_2NO_2$-$C2$, (c) $C_3N_2O_3$-$Imm2$, and (d) $C_3N_2O_3$-$Pca2_1$ crystals at 20 GPa.

To evaluate the energy density of the predicted compounds at ambient conditions, the ternary hull at 0 GPa is calculated, see Supplemental Figure S2. The crystals discussed above are at the following distance above the ternary convex hull at 0 GPa: $C_2N_2O$-Cmc2$_1$ - above 411 meV/atom; $C_2NO_2$-C2 - 550 meV/atom; $C_2NO_2$-C2/m - 532 meV/atom; $C_3N_2O_3$-Imm2 - 302 meV/atom; and $C_3N_2O_3$-Pca2$_1$ - 509 meV/atom. In comparison, the I2$_1$2$_1$2$_1$ crystal phase of the polymetic CO is 524 meV/atom above the convex hull. It is found that several of these crystals have a comparable energy density to that of the theoretical crystalline phase of carbon monoxide. Out of all the structures discussed, the $CN_2O$-P4$_3$[12] crystal is the furthest from the hull – above 968 meV/atom. The energy density of $CN_2O$-P4$_3$ was previously calculated to be 4.6 kJ/g[13]. The formation energy of $C_2N_2O$-Cmc2$_1$ is slightly lower than that of some other high energy density molecular crystals with the same stoichiometry[2–4], see Supplemental Table 1. However, the density of this crystal is larger and therefore its detonation performance is expected to be enhanced.

# Conclusion

The high pressure chemistry of $C_xN_yO_z$ ternary compounds has been investigated using first-principles crystal structure prediction methods in the range of pressures from 0 to 100 GPa. The ternary phase diagrams, i.e. ternary formation enthalpy/composition convex hulls, containing stable and metastable compounds have been constructed at several pressures based on the ternary formation enthalpy/composition convex hull. Only one thermodynamically stable compound is found on the ternary hull above 10 GPa with stoichiometry $C_2N_2O$ and space group Cmc2$_1$. Two other new metastable materials are found on the $(CO)_xN_y$ binary hull with stoichiometries $(CO)_2N$ and $(CO)_3N_2$. $(CO)_3N_2$ is on the binary hull from 10-100

GPa while $(CO)_2N$ is on the binary hull at 50 GPa up to 100 GPa. This indicates that these materials are energetically favored over isolated CO and nitrogen at high pressures. These structures are highly covalent extended crystals that consist of single bonded 3-fold coordinated nitrogen atoms, sp$^3$ bonded carbon atoms, and two-fold coordinated oxygen atoms. The only exception is $(CO)_3N_2$-Imm2 which is not as extended solid as the other materials are. Two new phases of $(CO)_3N_2$ compound are found with formation enthalpy lower than the molecular crystal carbonyll diisocynate synthesized previously, thus indicating the potential for synthesis of these materials in future experiments. The new materials have been characterized by their equations of state, Raman spectra, and chemical bonding.

## Supporting information

In the supporting information the results of DFT-MD simulations are presented to show the metastability of the ternary compounds. The ternary hull at ambient pressure is also given as it is used in the main text to estimate the ambient pressure energy densities of novel materials discovered in this work. The Supporting Information is available free of charge on the ACS Publications website at DOI: xxx

## Acknowledgement

The research is supported by the Defense Threat Reduction Agency, (grant No. HDTRA1-12- 1-0023) and Army Research Laboratory through Cooperative Agreement W911NF-16-2-0022. Simulations were performed using the NSF XSEDE supercomputers (grant No. TG-MCA08X040), DOE BNL CFN computational user facility, and USF Research Computing Cluster supported by NSF (grant No. CHE-1531590).

**Corresponding Authors**

email: oleynik@usf.edu

**Notes**

The authors declare no competing financial interests.

# References

(1) Klapotke, T. M.; Krumm, B.; Rest, S.; Scharf, R.; Schwabedissen, J.; Stammler, H.-G.; Mitzel, N. W. Carbonyl Diisocyanate CO(NCO) 2 : Synthesis and Structures in Solid State and Gas Phase. *J. Phys. Chem. A* **2016**, *120*, 4534–4541.

(2) Jones, P. G.; Ahrens, B.; Hopfner, T.; Hopf, H. 2,4,6-Tris(diazo)cyclohexane-1,3,5-trione. *Acta Cryst. C* **1997**, *53*, 783–786.

(3) Averkiev, B. B.; Antipin, M. Y.; Sheremetev, A. B.; Timofeeva, T. V. Four 3-cyanodifurazanyl ethers: Potential propellants. *Acta Cryst. C* **2003**, *59*, 383–387.

(4) Barbieux-Flammang, M.; Vandevoorde, S.; Flammang, R.; Wong, M. W.; Bibas, H.; Kennard, C. H. L.; Wentrup, C. Monomer, dimers and trimers of cyanogen N-oxide, N-C-C-N-O. an X-ray, FVT-MS/IR and theoretical investigation. *J. Chem. Soc. Perk. Trans.* **2000**, *2*, 473–478.

(5) Eremets, M. I.; Gavriliuk, A. G.; Trojan, I. A.; Dzivenko, D. A.; Boehler, R. Single-bonded cubic form of nitrogen. *Nat. Mat.* **2004**, *3*, 558–63.

(6) Tomasino, D.; Kim, M.; Smith, J.; Yoo, C. S. Pressure-induced Symmetry Lowering Transition in Dense Nitrogen to Layered Polymeric Nitrogen ( LP-N ) with Colossal Raman Intensity. *Phys. Rev. Lett.* **2014**, *113*, 205502.

(7) Iota, V.; Yoo, C.-S.; Klepeis, J.-H.; Jenei, Z.; Evans, W.; Cynn, H. Six-fold coordinated carbon dioxide VI. *Nat. Mat.* **2007**, *6*, 34–8.

(8) Shimizu, K.; Suhara, K.; Ikumo, M.; Eremets, M. I.; Amaya, K. Superconductivity in oxygen. *Nature* **1998**, *393*, 767–769.

(9) Lipp, M. J.; Evans, W. J.; Baer, B. J.; Yoo, C.-S. High-energy-density extended CO solid. *Nat. Mat.* **2005**, *4*, 211–5.

(10) Christe, K. Recent Advances in the Chemistry of N5+, N5- and High-Oxygen Compounds. *Propellants, Explos , Pyrotech* **2007**, *32*, 194–204.

(11) Mailhiot, C.; Yang, L.; McMahan, A. Polymeric nitrogen. *Phys. Rev. B* **1992**, *46*, 14419–14435.

(12) Raza, Z.; Pickard, C.; Pinilla, C.; Saitta, A. High Energy Density Mixed Polymeric Phase from Carbon Monoxide and Nitrogen. *Phys. Rev. Lett.* **2013**, *111*, 235501.

(13) Zhu, C.; Li, Q.; Zhou, Y.; Zhang, M.; Zhang, S.; Li, Q. Exploring High-Pressure Structures of $N_2CO$. *J. Phys. Chem. C* **2014**, *118*, 27252–27257.

(14) Steele, B. A.; Oleynik, I. I. Sodium pentazolate: A nitrogen rich high energy density material. *Chem. Phys. Lett.* **2016**, *643*, 21–26.

(15) Steele, B. A.; Stavrou, E.; Crowhurst, J. C.; Zaug, J. M.; Prakapenka, V. B.; Oleynik, I. I. High-Pressure Synthesis of a Pentazolate Salt. *Chem. Mater.* **2017**, *29*, 735–741.

(16) Steele, B. A.; Oleynik, I. I. Pentazole and Ammonium Pentazolate: Crystalline Hydro-Nitrogens at High Pressure. *J. Phys. Chem. A* **2017**, *121*, 1808–1813.

(17) Shen, Y.; Oganov, A. R.; Qian, G.; Zhang, J.; Dong, H.; Zhu, Q.; Zhou, Z. Novel lithium-nitrogen compounds at ambient and high pressures. *Sci. Rep.* **2015**, *5*, 14204.

(18) Peng, F.; Yao, Y.; Liu, H.; Ma, Y. Crystalline LiN5 Predicted from First-Principles as a Possible High-Energy Material. *J. Phys. Chem. Lett.* **2015**, *6*, 2363–6.

(19) Peng, F.; Han, Y.; Liu, H.; Yao, Y. Exotic stable cesium polynitrides at high pressure. *Sci. Rep.* **2015**, *5*, 16902.

(20) Glass, C. W.; Oganov, A. R.; Hansen, N. USPEX-evolutionary crystal structure prediction. *Comp. Phys. Comm.* **2006**, *175*, 713–720.

(21) Lyakhov, A. O.; Oganov, A. R.; Valle, M. How to predict very large and complex crystal structures. *Comp. Phys. Comm.* **2010**, *181*, 1623–1632.

(22) Oganov, A. R.; Glass, C. W. Crystal structure prediction using ab initio evolutionary techniques: principles and applications. *J. Chem. Phys.* **2006**, *124*, 244704.

(23) Gregoryanz, E.; Goncharov, A. F.; Sanloup, C.; Somayazulu, M.; Mao, H.-k.; Hemley, R. J. High P-T transformations of nitrogen to 170 GPa. *J. Chem. Phys.* **2007**, *126*, 184505.

(24) Desgreniers, S.; Vohra, Y. K.; Ruoff, A. L. Optical response of very high density solid oxygen to 132 GPa. *J. Phys. Chem.* **1990**, *94*, 1117–1122.

(25) Grumbach, M.; Martin, R. Phase diagram of carbon at high pressures and temperatures. *Phys. Rev. B* **1996**, *54*, 15730–15741.

(26) Li, D.; Oganov, A. R.; Dong, X.; Zhou, X.-F.; Zhu, Q.; Qian, G.; Dong, H. Nitrogen oxides under pressure: stability, ionization, polymerization, and superconductivity. *Sci Rep* **2015**, *5*, 16311.

(27) Sun, J.; Klug, D. D.; Martonak, R.; Montoya, J. A.; Lee, M.-S.; Scandolo, S.; Tosatti, E. High-pressure polymeric phases of carbon dioxide. *PNAS* **2009**, *106*, 6077–6081.

(28) Dong, H.; Oganov, A. R.; Zhu, Q.; Qian, G.-R. The phase diagram and hardness of carbon nitrides. *Sci. Rep.* **2015**, *5*, 9870.

(29) Perdew, J.; Burke, K.; Ernzerhof, M. Generalized Gradient Approximation Made Simple. *Phys. Rev. Lett.* **1996**, *77*, 3865–3868.

(30) Kresse, G.; Furthmiiller, J. Efficiency of ab-initio total energy calculations for metals and semiconductors using a plane-wave basis set. *Comp. Mat. Sci.* **1996**, *6*, 15–50.

(31) Kresse, G.; Joubert, D. From ultrasoft pseudopotentials to the projector augmented-wave method. *Phys. Rev. B* **1999**, *59*, 1758–1775.

(32) Delley, B. From molecules to solids with the DMol3 approach. *J. Chem. Phys.* **2000**, *113*, 7756.

(33) Porezag, D.; Pederson, M. R. Infrared intensities and Raman-scattering activities within density-functional theory. *Phys. Rev. B* **1996**, *54*, 7830–7836.

(34) Ivanovskii, A. L.; Medvedeva, N. I.; Kontsevoi, O. Y.; Shveikin, G. P. Ab-initio Calculations of the Electronic Structure and Cohesive Properties of the Orthorhombic Oxynitrides. *Phys. Stat. Sol. b* **2000**, *221*, 647.

(35) Xiong, Q.-Y.; Shen, Q.-X.; Li, R.-Z.; Shen, J.; Tian, F.-Y. First-principle investigation on the thermodynamics of $X_2N_2O$ ( X = C, Si, Ge) compounds. *Chinese Physics B* **2016**, *25*, 026501.

(36) Hocking, W. H.; Gerry, M. C. L. Microwave Spectrum and Structure of Cyanogen Isocyanate. *J. C. S. Chem. Comm.* **1973**, 47.

(37) Schmidt, C. L.; Jansen, M. Cyanogen isocyanate (NC-NCO) revisited: thermal and chemical reactivity of a hydrogen-free precursor to C-N-O polymers. *J. Mat. Chem.* **2010**, *20*, 110–116.

(38) Sun, J.; Klug, D. D.; Pickard, C. J.; Needs, R. J. Controlling the Bonding and Band Gaps of Solid Carbon Monoxide with Pressure. *Phys. Rev. Lett.* **2011**, *106*, 145502.

(39) Ciezak-Jenkins, J. A.; Steele, B. A.; Borstad, G. M.; Oleynik, I. I. Structural and spectroscopic studies of nitrogen-carbon monoxide mixtures: Photochemical response and observation of a novel phase. *J. Chem. Phys.* **2017**, *146*, 184309.

(40) Ceppatelli, M.; Pagliai, M.; Bini, R.; Jodl, H. J. High-Pressure Photoinduced Synthesis of Polynitrogen in delta and epsilon Nitrogen Crystals Substitutionally Doped with CO. *J Phys Chem C* **2015**, *119*, 130–140.

(41) Zeng, X.; Gerken, M.; Beckers, H.; Willner, H. Synthesis and characterization of carbonyl diazide, OC(N3)2. *Inorg Chem* **2010**, *49*, 9694–9.

(42) Pauling, L. *General Chemistry*; Dover Publication: New York, 1988; p 192.

(43) de Jong, M.; Chen, W.; Angsten, T.; Jain, A.; Notestine, R.; Gamst, A.; Sluiter, M.; Krishna Ande, C.; van der Zwaag, S.; Plata, J. J.; Toher, C.; Curtarolo, S.; Ceder, G.; Persson, K. A.; Asta, M. Charting the complete elastic properties of inorganic crystalline compounds. *Sci. Data* **2015**, *2*, 150009.

(44) Anderson, O. L.; Nafe, J. E. The bulk modulus-volume relationship for oxide compounds and related geophysical problems. *J. Geophys. Res.* **1965**, *70*, 3951–3963.

# Graphical TOC Entry

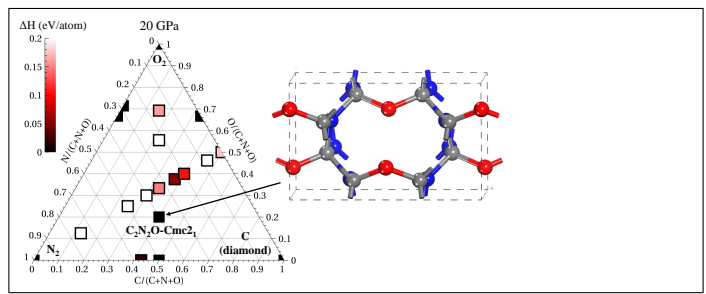

For Table of Contents Only: The ternary carbon, nitrogen, oxygen convex hull at 20 GPa: each square represents a crystal structure and the color represents the distance to the three-dimensional convex hull. The crystal structure of $C_2N_2O$ - newly discovered material, which is the only thermodynamically stable crystal at 20 GPa, is shown as well.